\documentclass{elsart5p}

\usepackage{graphics}
\usepackage{graphicx}
\usepackage{amssymb}
\begin{document}

\begin{frontmatter}



\title{Polarized neutron scattering study on antiferromagnetic states in CeRh$_{0.6}$Co$_{0.4}$In$_5$}

\author[IBARAKI]{M. Yokoyama},
\ead{makotti@mx.ibaraki.ac.jp}
\author[IBARAKI]{N. Oyama},
\author[HOKUDAI]{H. Amitsuka},
\author[HOKUDAI]{S. Oinuma},
\author[HOKUDAI]{I. Kawasaki},
\author[HOKUDAI]{K. Tenya},
\author[ISSP]{M. Matsuura},
\author[ISSP]{K. Hirota}

\address[IBARAKI]{Faculty of Science, Ibaraki University, Mito 310-8512, Japan}
\address[HOKUDAI]{Graduate School of Science, Hokkaido University, Sapporo 060-0810, Japan}
\address[ISSP]{Neutron Science Laboratory, Institute for Solid State Physics, The University of Tokyo, Tokai 319-1106, Japan}

\begin{abstract}
Polarized neutron scattering experiments were performed on mixed compound CeRh$_{0.6}$Co$_{0.4}$In$_5$ to clarify the nature of the low-temperature ordered states. Three nonequivalent Bragg peaks, characterized by the wave vectors of $q_{\rm h} \sim (1/2,1/2,0.3)$, $q_1 \sim (1/2,1/2,0.4)$ and $q_{\rm c}=(1/2,1/2,1/2)$, were observed at 1.4 K. These Bragg peaks are found to occur entirely in spin-flip channel. This indicates that these Bragg peaks originate from the magnetic scattering, {\it i.e.}, the antiferromagnetic orders with three different modulations appear in this compound.
\end{abstract}


\end{frontmatter}

% main text
Quantum critical behavior observed in the heavy-fermion compounds has been attracting much interest in recent years. The ternary tetragonal compound CeRhIn$_5$ shows an incommensurate antiferromagnetic (I-AF) order with the modulation of $q=(1/2,1/2,0.297)$ below $T_{\rm N}=3.8\ {\rm K}$ \cite{rf:Hegger2000,rf:Bao2000}. It is found in the mixed compounds CeRh$_{1-x}$Co$_x$In$_5$ \cite{rf:Zapf2001,rf:Jeffries2005} that $T_{\rm N}$ of the I-AF order is weakly reduced  by substituting Co for Rh, and then approaches zero at $x_c\sim 0.7$. At the same time,  the superconducting (SC) phase appears above $x = 0.4$.
Recently, our elastic neutron scattering experiments revealed that a commensurate antiferromagnetic (C-AF) order with the propagation vector of $q_{\rm c}=(1/2,1/2,1/2)$ as well as the I-AF order with $q_{\rm h}=(1/2,1/2,0.306)$ evolve at $x=0.4$ \cite{rf:Yoko2006}. We have also observed weak Bragg peaks characterized by the incommensurate wave vector $q_1=(1/2,1/2,0.403)$, but their origin is still unclear. To clarify the nature of these Bragg peaks, we have performed polarized neutron scattering experiments for CeRh$_{0.6}$Co$_{0.4}$In$_5$.

Single crystals of CeRh$_{1-x}$Co$_{x}$In$_5$ with $x=0.4$ were grown by the In-flux method. To minimize the effects of neutron absorption by Rh and In, we prepared a rod-type sample with the dimensions of $\sim 1.8\times 1.8\times 15\ {\rm mm}^3$ by means of a spark erosion. The electron probe microanalysis (EPMA) measurements for the sample indicate that the homogeneous distributions of the elements over the averaged area ($\sim 1\ \mu$m$^2$) are achieved, but the estimated Co/Rh concentration ($x\sim 0.5$) is slightly larger than the nominal $x$ value (0.4) \cite{rf:Ami2007}. Similar deviation of $x$ is also observed in the sample investigated previously \cite{rf:Yoko2006}. Despite these results, we will use the nominal concentration $x$ throughout this article for simplicity and clarity. 

The polarized neutron scattering experiments were carried out on the triple-axis spectrometer ISSP-PONTA (5G) located at the research reactor JRR-3M of the JAEA, Tokai, Japan. The neutron momentum $k=4.05\ {\rm \AA}^{-1}$ was chosen by monochromator and analyzer (Heusler), and a combination of B-80'-80'-B collimators and a pyrolytic graphite (PG) filter was used. The spin direction of the incident neutron beam was tuned to be parallel to the scattering vector $Q$ in Helmholtz coils. In the experimental geometry presently used, it is expected that the nuclear scattering is completely non-spin flip, while any magnetic scattering flips the neutron spins. The polarization ratio of the neutron spins estimated from the nuclear Bragg-peak intensities was $\sim 15$. We also performed the unpolarized neutron scattering experiments on the triple-axis spectrometer ISSP-GPTAS (4G). We selected the neutron momentum $k=3.79\ {\rm \AA}^{-1}$ by PG monochromator and analyzer, and used a 40'-40'-40'-80' collimator and two PG filters. Both the measurements were performed in the $(hhl)$ scattering planes and the temperature range of 1.4--10 K.

Fig.\ 1 shows neutron scattering patterns in spin-flip (SF) and non-spin-flip (NSF) scattering channels at 1.4 K, obtained from the $Q=(1/2,1/2,1+\zeta)$ $(0.46\le\zeta \le0.74)$ scans in the polarized neutron scattering measurements. The profiles of the (110) nuclear Bragg peak are also displayed in the inset of Fig.\ 1. We have observed three nonequivalent Bragg peaks at $(1/2,1/2,3/2)$, $(1/2,1/2,1.60(1))$, and $(1/2,1/2,1.697(5))$ in the SF scattering channel, whose positions correspond with those expected from the ordering vectors $q_{\rm c}$, $q_1$ and $q_{\rm h}$, respectively. The width of the Bragg peak at (1/2,1/2,1.60) is clearly larger than others, indicating that the order with the modulation of  $q_1$ occurs in a shorter range along the $c$ axis.  This feature is also observed in our previous measurements \cite{rf:Yoko2006}. On the other hand, no significant reflection is detected in the NSF scattering channel within the experimental accuracy. These results indicate that all these Bragg peaks originate from the magnetic scattering, {\it i.e.}, the antiferromagnetic (AF) orders with the modulations of $q_{\rm c}$, $q_1$ and $q_{\rm h}$ occur at $x\sim 0.4$.  
\begin{figure}[tbp]
\begin{center}
\includegraphics[keepaspectratio,width=0.48\textwidth]{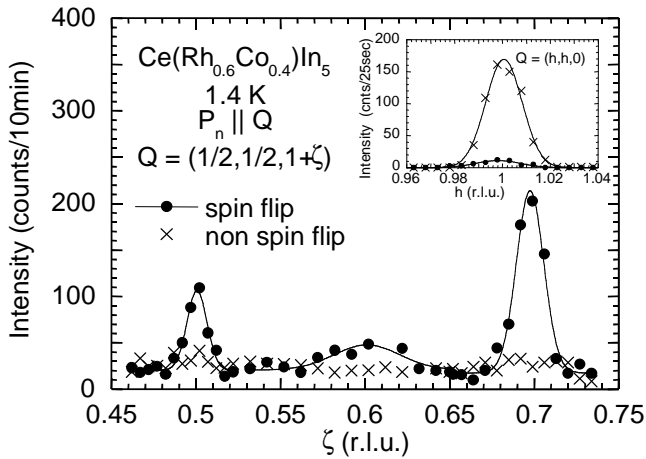}
\end{center}
\caption{Neutron scattering patterns in spin-flip and non-spin-flip scattering channels obtained by the scans at $Q=(1/2,1/2,1+\zeta)$ $(0.46\le\zeta \le0.74)$ at 1.4 K for CeRh$_{0.6}$Co$_{0.4}$In$_5$. The inset shows the profiles of the (110) nuclear Bragg peak.}
\end{figure}

We have estimated the magnitudes of the volume-averaged ordered moments $\mu_{\rm h},\ \mu_1$ and $\mu_{\rm c}$ for the AF orders with the modulations of $q_{\rm h}$, $q_1$ and $q_{\rm c}$, respectively, from the integrated intensities of the magnetic Bragg peaks obtained by longitudinal and transverse scans with the unpolarized neutron scattering technique. In accordance with the previous reports \cite{rf:Bao2000,rf:Yoko2006,rf:Curro2000,rf:Majumdar2002}, we assume that both the I-AF orders have the helical AF structures, and the ordered moments in all the AF states lie in the tetragonal basal plane. In Table 1, we show the $\mu_{\rm h},\ \mu_1$ and $\mu_{\rm c}$ values at 1.4 K for the sample (labeled as \#2) presently investigated, together with those for previous one (\#1) \cite{rf:Yoko2006}. The $\mu_{\rm c}/\mu_{\rm h}$ and $\mu_{\rm 1}/\mu_{\rm h}$ values for \#2 are found to be significantly smaller than those for \#1, while the total moments for \#1 and \#2 defined by $\sqrt{\mu_{\rm h}^2+\mu_1^2+\mu_{\rm c}^2}$ are in good agreement with each other. This suggests that the fraction of the moment distributed to each AF state is very sensitive to the sample conditions, such as Co concentration and crystal quality. We consider that the I-AF order with $q_1$ is generated in parts of the sample because of the large AF Bragg-peak width.  However, it is not clear in the present stage whether the AF states with $q_{\rm h}$ and $q_{\rm c}$ are microscopically coexistent or inhomogeneously separated. This issue may be resolved by performing NMR and $\mu$SR experiments in the future.
\begin{table}[tbp]
\caption{The volume-averaged ordered moments of the AF orders with the modulations of $q_{\rm h}$, $q_1$ and $q_{\rm c}$ at 1.4 K for CeRh$_{0.6}$Co$_{0.4}$In$_5$. $\mu_{\rm T}$ indicates the total moment defined by $\mu_{\rm T}=\sqrt{\mu_{\rm h}^2+\mu_1^2+\mu_{\rm c}^2}$.}
\begin{center}
\begin{tabular}{ccccccccccccc}\hline 
sample && $\mu_{\rm h}$($\mu_{\rm B}$) && $\mu_1$($\mu_{\rm B}$) && $\mu_{\rm c}$($\mu_{\rm B}$) && $\mu_{\rm 1}/\mu_{\rm h}$ && $\mu_{\rm c}/\mu_{\rm h}$ &&  $\mu_{\rm T}$($\mu_{\rm B}$) \\\hline
\#1 && 0.38(3) && 0.17(2) && 0.21(2) && 0.45(2) && 0.55(1) && 0.47(5) \\
\#2 && 0.41(2) && 0.15(2) && 0.18(1) && 0.37(2) && 0.44(1) &&  0.47(4)\\
\hline 
\end{tabular}
\end{center}
\end{table}

We have observed new AF states with the modulations of $q_{\rm c}$ and $q_{1}$ as well as the I-AF state with $q_{\rm h}$ in CeRh$_{0.6}$Co$_{0.4}$In$_5$. The evolution of the C-AF state near the superconducting phase is also found in recent neutron scattering experiments for Ir-doped CeRhIn$_5$ \cite{rf:Christianson2005}, Cd-doped CeCoIn$_5$ \cite{rf:Nicklas2007}, and NQR experiments for pure CeRhIn$_5$ under pressure \cite{rf:Yashima2007}, implying that the C-AF correlation plays a key role in the stability of the SC phase in the CeMIn$_5$ systems (M=Co, Rh and Ir). In addition, the I-AF order with the modulation of $\sim q_{\rm 1}$ is also detected by the high-pressure neutron scattering measurements for pure CeRhIn$_5$ \cite{rf:Majumdar2002}. The evolutions of the AF states with a variety of the $c$-axis modulations may be attributed to two dimensional characteristics of these compounds.

In summary, our polarized neutron scattering experiments on CeRh$_{0.6}$Co$_{0.4}$In$_5$ revealed that all of the Bragg-peaks associated with the incommensurate ($q_{\rm h}$ and $q_{\rm 1}$) and commensurate ($q_{\rm c}$) structures occur entirely due to the magnetic scattering. This indicates that the AF orders with three different modulations appear at $x\sim 0.4$.

This work was supported by a Grant-in-Aid for Scientific Research from the Ministry of Education, Culture, Sports, Science and Technology of Japan.

\end{document}